\documentclass[a4paper,11pt]{article}%
\usepackage{jheppub}
\usepackage{amsmath}
\usepackage{amsfonts}
\usepackage{amssymb}%
\usepackage{graphicx}
\providecommand{\U}[1]{\protect\rule{.1in}{.1in}}
\affiliation{Budker Institute of Nuclear Physics and Novosibirsk State University,\\630090 Novosibirsk, Russia}
\emailAdd{A.V.Grabovsky@inp.nsk.su}
\abstract{It is demonstrated that the NLO forward BFKL equation can be solved in the space of its Born eigenfunctions.}
\arxivnumber{}
\notoc
\begin{document}

\title{\boldmath On the solution to the NLO forward BFKL equation}
\author{A. V. Grabovsky}
\maketitle

\flushbottom

\section{Introduction}

The Balitsky-Fadin-Kuraev-Lipatov (BFKL) equation \cite{Fadin:1975cb}%
--\cite{Balitsky:1978ic} governs the energy evolution of the pomeron Green
function. The eigenfunctions of the Born BFKL kernel were found in
\cite{Lipatov:1985uk}, which allows one to solve the leading order (LO)
equation. The next to leading order (NLO) corrections to the forward kernel of
the BFKL equation were calculated in \cite{Fadin:1998py}. However, due to the
running of the coupling constant the Born eigenfunctions are not the
eigenfunctions of the NLO equation. The NLO forward BFKL equation and
different approaches to its solution allowing for the running coupling effects
were studied extensively, see \cite{Kovchegov:1998ae}, lectures
\cite{Salam:1999cn} and references therein. In recent paper
\cite{Chirilli:2013kca} there was constructed an eigenfunction of the forward
NLO BFKL kernel and it was used to build a solution to the forward NLO BFKL
equation. Here it is shown that one can solve the forward BFKL equation in the
space of the Born eigenfunctions without constructing the NLO eigenfunction.

\section{The solution of the forward NLO BFKL equation}

Our notation is the same as in \cite{Fadin:2009za}. Thus the Reggeon
transverse momenta (and the conjugate coordinates) in initial and final
$t$-channel states are $\vec{q}_{i}^{\;\prime}(\vec{r}_{i}^{\;\prime})$ and
$\vec{q}_{i}$ ($\vec{r}_{i}),$ $i=1,2$. The state normalization is
\begin{equation}
\langle\vec{q}|\vec{q}^{\;\prime}\rangle=\delta(\vec{q}-\vec{q}^{\;\prime
})\;,\;\;\;\;\;\langle\vec{r}|\vec{r}^{\;\prime}\rangle=\delta(\vec{r}-\vec
{r}^{\;\prime})\;,\quad\quad\langle\vec{r}|\vec{q}\rangle=\frac{e^{i\vec
{q}\,\vec{r}}}{2\pi}.\;\label{normalization}%
\end{equation}
The $s$-channel discontinuities of scattering amplitudes for the processes
$A+B\rightarrow A^{\prime}+B^{\prime}$ have the form
\begin{equation}
-4i(2\pi)^{2}\delta(\vec{p}_{A}^{\,\,\prime}+\vec{p}_{B}^{\,\,\prime}-\vec
{p}_{A}-\vec{p}_{B})\mbox{disc}_{s}\mathcal{A}_{AB}^{A^{\prime}B^{\prime}%
}=\langle A^{\prime}\bar{A}|\hat{G}\frac{1}{\hat{\vec{q}}_{1}^{\;2}\hat
{\vec{q}}_{2}^{\;2}}|\bar{B}^{\prime}B\rangle
\;.\label{discontinuity representation}%
\end{equation}
In this expression $\hat{G}$ is the BFKL Green function, obeying the BFKL
equation. This equation can be written in the operator form as
\begin{equation}
\frac{\partial\hat{G}}{\partial Y}=\hat{K}\hat{G},\,\quad\hat{G}|_{Y=Y_{0}%
}=\hat{1},
\end{equation}
where $Y$ is the rapidity and $Y_{0}$ is given by an appropriate energy scale
set by the impact factors. For the forward scattering, we define the matrix
element in the coordinate representation as
\begin{equation}
\langle\vec{r}|\hat{K}|\vec{r}^{\,\,\,\prime}\rangle=\int\langle\vec{r}%
_{1}\vec{r}_{2}|\hat{K}|\vec{r}_{1}^{\;\prime}\vec{r}_{2}^{\;\prime}%
\rangle\delta(\vec{r}_{1^{\prime}2^{\prime}}-\vec{r}^{\,\,\prime})d^{2}{r}%
_{1}^{\;\prime}d^{2}{r}_{2}^{\;\prime}~,\quad\label{forward_kernel_coord}%
\end{equation}
with $\vec{r}=\vec{r}_{12}$. Hence%
\begin{equation}
\langle\vec{q}_{1}|\hat{K}|\vec{q}_{1}^{\,\,\prime}\rangle=\int\frac{d\vec
{x}d\vec{z}}{\left(  2\pi\right)  ^{2}}e^{-i\left[  \vec{q}_{1}\vec{r}-\vec
{q}_{1}^{\,\,\prime}\vec{r}^{\,\,\prime}\right]  }\langle\vec{r}|\hat{K}%
|\vec{r}^{\,\,\prime}\rangle.\label{forward_Fourier}%
\end{equation}
Then discontinuity (\ref{discontinuity representation}) turns into
\begin{equation}
-4i(2\pi)^{2}\mbox{disc}_{s}\mathcal{A}_{AB}^{AB}=\langle A\bar{A}|\hat
{G}\,\,\hat{\vec{q}}_{1}^{\;-4}|\bar{B}B\rangle
\;.\label{forward discontinuity representation}%
\end{equation}
We will work in the space of the Born kernel eigenfunctions $|n\nu\rangle$%
\begin{equation}
\langle\vec{r}|n\nu\rangle=\frac{1}{\pi\sqrt{2}}e^{in\phi}(\vec{r}%
^{\,\,2})^{-\frac{1}{2}+i\nu},\quad\langle\vec{r}|n\nu\rangle\langle n\nu
|\vec{r}^{\,\,\prime}\rangle=\delta(\vec{r}-\vec{r}^{\,\,\prime}),\quad\langle
n\nu|\vec{r}\rangle\langle\vec{r}|m\sigma\rangle=\delta_{nm}\delta(\nu
-\sigma).
\end{equation}%
\begin{equation}
\langle\vec{r}|n\nu\rangle=\frac{1}{\pi\sqrt{2}}e^{in\phi}(\vec{r}%
^{\,\,2})^{-\frac{1}{2}+i\nu}=\frac{1}{\pi\sqrt{2}}e^{in\phi}(\vec{r}%
^{\,\,2})^{\gamma},\quad\gamma=-\frac{1}{2}+i\nu.
\end{equation}
Hereafter the summation and integration over repeated indices is performed. In
this space the discontinuity is proportional to the matrix element%
\begin{equation}
-4i(2\pi)^{2}\mbox{disc}_{s}\mathcal{A}_{AB}^{AB}=\langle A\bar{A}|n\nu
\rangle\langle n\nu|\hat{G}|m\sigma\rangle\langle m\sigma|\hat{\vec{q}}%
_{1}^{\;-4}|\bar{B}B\rangle\;.
\end{equation}
and the BFKL equation has the form%
\begin{equation}
\frac{\partial\langle n\nu|\hat{G}|m\sigma\rangle}{\partial Y}=\langle
n\nu|\hat{K}|h\rho\rangle\langle h\rho|\hat{G}|m\sigma\rangle,\quad\langle
n\nu|\hat{G}|m\sigma\rangle|_{Y=Y_{0}}=\delta_{nm}\delta(\nu-\sigma).
\end{equation}
Also the impact factors have the form%
\begin{equation}
\Phi_{A}\left(  n,\nu\right)  =\langle A^{\prime}\bar{A}|n\nu\rangle,\quad
\Phi_{B}\left(  m,\sigma\right)  =\langle m\sigma|\hat{\vec{q}}_{1}%
^{\;-4}|\bar{B}^{\prime}B\rangle,
\end{equation}
while the kernel reads (see (52)-(55) in \cite{Fadin:2009za})%
\begin{equation}
\langle n\nu|\hat{K}_{M}|h\rho\rangle=\left[  \bar{\alpha}\chi(n,\rho
)+\frac{\bar{\alpha}^{2}}{4}\delta(n,\rho)\right]  \delta_{nh}\delta\left(
\rho-\nu\right)  -i\frac{\bar{\alpha}^{2}\beta}{4}\chi(n,\rho)\delta
_{nh}\delta^{\prime}\left(  \rho-\nu\right)  ,\,
\end{equation}%
\begin{equation}
\beta=\frac{11}{3},\quad\bar{\alpha}=\frac{\,\alpha_{s}(\mu^{2})N_{c}\,}{\pi
},\quad\chi(n,\nu)=2\psi(1)-\psi(\frac{1+n}{2}+i\nu)-\psi(\frac{1+n}{2}%
-i\nu)\,,\label{l12}%
\end{equation}
and $\delta(n,\nu)$ contains all the terms from (53) in \cite{Fadin:2009za}
without the derivative. For convenience the Moebius kernel defined in the
coordinate representation is used here. However, the same procedure may be
done with the original kernel obtained in the momentum space in
\cite{Fadin:1998py}. At once we can perform the convolution of the Green
function and the second impact factor. As a result for the equation we obtain
\[
\frac{\partial\langle n\nu|\hat{G}\,\hat{\vec{q}}_{1}^{\,-4}|\bar{B}B\rangle
}{\partial Y}-i\frac{\bar{\alpha}^{2}\beta}{4}\frac{\partial\left(  \chi
(n,\nu)\langle n\nu|\hat{G}\,\hat{\vec{q}}_{1}^{\,-4}|\bar{B}B\rangle\right)
}{\partial\nu}%
\]%
\begin{equation}
=\left[  \bar{\alpha}\chi(n,\nu)+\frac{\bar{\alpha}^{2}}{4}\delta
(n,\nu)\right]  \langle n\nu|\hat{G}\,\hat{\vec{q}}_{1}^{\,\,\,-4}|\bar
{B}B\rangle\label{equation}%
\end{equation}
with the initial condition
\begin{equation}
\langle n\nu|\hat{G}\,\hat{\vec{q}}_{1}^{\,-4}|\bar{B}B\rangle|_{Y=Y_{0}}%
=\Phi_{B}\left(  n,\nu\right)  .\label{initial_condition}%
\end{equation}
For simplicity we will write $\chi(n,\nu)\langle n\nu|\hat{G}\,\hat{\vec{q}%
}^{\,\,-4}|\bar{B}^{\prime}B\rangle=\mathbf{G}.$ In this notation the equation
is%
\begin{equation}
\frac{\partial\mathbf{G}}{\partial Y}-i\frac{\bar{\alpha}^{2}\beta}{4}%
\chi(n,\nu)\frac{\partial\mathbf{G}}{\partial\nu}=\left[  \bar{\alpha}%
\chi(n,\nu)+\frac{\bar{\alpha}^{2}}{4}\delta(n,\nu)\right]  \mathbf{G,\quad}%
\end{equation}%
\begin{equation}
\mathbf{G}_{0}\left(  n,\nu\right)  =\mathbf{G}|_{Y=Y_{0}}=\chi(n,\nu)\Phi
_{B}\left(  n,\nu\right)  .
\end{equation}
Its two first integrals are
\begin{equation}
c_{1}=Y-Y_{0}-\frac{4i}{\bar{\alpha}^{2}\beta}\int_{\nu_{0}}^{\nu}\frac
{dl}{\chi\left(  n,l\right)  },\quad c_{2}=\mathbf{G}e^{-i\left[  \frac
{4}{\bar{\alpha}\beta}(\nu-\nu_{0})+\int_{\nu_{0}}^{\nu}\frac{\delta\left(
n,l\right)  }{\chi\left(  n,l\right)  \beta}dl\right]  }.
\end{equation}
The general solution to this equation depends on an arbitrary function $f$%
\begin{equation}
\mathbf{G}=e^{i\left[  \frac{4}{\bar{\alpha}\beta}(\nu-\nu_{0})+\int_{\nu_{0}%
}^{\nu}\frac{\delta\left(  n,l\right)  }{\chi\left(  n,l\right)  \beta
}dl\right]  }f\left(  \frac{\bar{\alpha}^{2}\beta}{4}i(Y-Y_{0})+\int_{\nu_{0}%
}^{\nu}\frac{dl}{\chi\left(  n,l\right)  }\right)  .\label{gensol}%
\end{equation}
Using initial condition (\ref{initial_condition}), we can find $f$ from the
equation
\begin{equation}
\mathbf{G}_{0}\left(  n,\nu\right)  =e^{i\left[  \frac{4}{\bar{\alpha}\beta
}(\nu-\nu_{0})+\int_{\nu_{0}}^{\nu}\frac{\delta\left(  n,l\right)  }%
{\chi\left(  n,l\right)  \beta}dl\right]  }f\left(  \int_{\nu_{0}}^{\nu}%
\frac{dl}{\chi\left(  n,l\right)  }\right)  .
\end{equation}
To solve it we introduce
\begin{equation}
F(\nu)=\int_{\nu_{0}}^{\nu}\frac{dl}{\chi\left(  n,l\right)  }.
\end{equation}
Hence%
\begin{equation}
f\left(  t\right)  =\mathbf{G}_{0}(n,F^{-1}(t))\exp\left[  \frac{-4i}%
{\bar{\alpha}\beta}(F^{-1}(t)-\nu_{0})-\frac{i}{\beta}\int_{\nu_{0}}%
^{F^{-1}\left(  t\right)  }\frac{\delta\left(  n,l\right)  }{\chi\left(
n,l\right)  }dl\right]  .
\end{equation}
Then the formal general solution reads%
\[
\mathbf{G}=\mathbf{G}_{0}(n,F^{-1}(F\left(  v\right)  +\frac{\bar{\alpha}%
^{2}\beta i(Y-Y_{0})}{4}))
\]%
\begin{equation}
\times e^{-\frac{4i}{\bar{\alpha}\beta}(F^{-1}\left(  F(\nu)+\frac{\bar
{\alpha}^{2}\beta i}{4}(Y-Y_{0})\right)  -\nu)}e^{-\frac{i}{\beta}\int_{\nu
}^{F^{-1}\left(  F(\nu)+\frac{\bar{\alpha}^{2}\beta i}{4}(Y-Y_{0})\right)
}\frac{\delta\left(  n,l\right)  }{\chi\left(  n,l\right)  }dl}.
\end{equation}%
\[
\langle n\nu|\hat{G}\,\,\hat{\vec{q}}_{1}^{\,-4}|\bar{B}^{\prime}B\rangle
=\Phi_{B}(n,F^{-1}\left(  F(\nu)+\frac{\bar{\alpha}^{2}\beta i}{4}%
(Y-Y_{0})\right)  )\frac{\chi(n,F^{-1}\left(  F(\nu)+\frac{\bar{\alpha}%
^{2}\beta i}{4}(Y-Y_{0})\right)  )}{\chi(n,\nu)}\,
\]%
\begin{equation}
\times e^{-\frac{4i}{\bar{\alpha}\beta}(F^{-1}\left(  F(\nu)+\frac{\bar
{\alpha}^{2}\beta i}{4}(Y-Y_{0})\right)  -\nu)}e^{-\frac{i}{\beta}\int_{\nu
}^{F^{-1}\left(  F(\nu)+\frac{\bar{\alpha}^{2}\beta i}{4}(Y-Y_{0})\right)
}\frac{\delta\left(  n,l\right)  }{\chi\left(  n,l\right)  }dl}%
.\label{formalSolution}%
\end{equation}
Here we can keep only the terms up to $\bar{\alpha}^{m+1}Y^{m}$ order since we
work in NLO. With this accuracy in the decomposition of $F^{-1}\left(
F(\nu)+\frac{\bar{\alpha}^{2}\beta}{4i}(Y-Y_{0})\right)  $ at $F\left(
\nu\right)  $ we can take only 3 first terms%
\begin{equation}
F^{-1}\left(  F(\nu)+\frac{\bar{\alpha}^{2}\beta i}{4}(Y-Y_{0})\right)
\simeq\nu+\chi\left(  n,\nu\right)  \frac{\bar{\alpha}^{2}\beta i}{4}%
(Y-Y_{0})+\frac{\chi\left(  n,\nu\right)  \chi^{\prime}\left(  n,\nu\right)
}{2}\left(  \frac{\bar{\alpha}^{2}\beta i}{4}Y\right)  ^{2}%
,\label{F-1decomposition}%
\end{equation}
since%
\begin{equation}
\frac{dF^{-1}(x)}{dx}|_{x=F\left(  t\right)  }=\chi\left(  n,t\right)
,\quad\frac{d^{2}F^{-1}(x)}{d^{2}x}|_{x=F\left(  t\right)  }=\chi\left(
n,t\right)  \chi^{\prime}\left(  n,t\right)  .
\end{equation}
Therefore with NLO accuracy%
\[
\langle n\nu|\hat{G}\,\hat{\vec{q}}_{1}^{\,\,-4}|\bar{B}B\rangle=\left(
\Phi_{B}(n,\nu)+\frac{\bar{\alpha}^{2}\beta i}{4}[\Phi_{B}(n,\nu)\chi\left(
n,\nu\right)  ]^{\prime}Y\right)
\]%
\begin{equation}
\times e^{\bar{\alpha}\chi\left(  n,\nu\right)  (Y-Y_{0})+\frac{\bar{\alpha
}^{2}}{4}\delta\left(  n,\nu\right)  Y+\frac{\bar{\alpha}^{3}\beta i}{8}%
\chi\left(  n,\nu\right)  \chi^{\prime}\left(  n,\nu\right)  Y^{2}%
}.\label{<nu|BB>}%
\end{equation}
During the derivation there were several subtle points, e.g. the treatment of
$F(\nu)$'s singularities. They may be addressed. However it is easier to plug
the solution into (\ref{equation}) and check that it satisfies the equation up
to terms proportional to $\bar{\alpha}^{n+2}Y^{n}$ including them. Then the
contribution to discontinuity (\ref{forward discontinuity representation})
will have the form, which is the main result of this paper%
\[
\langle A\bar{A}|\hat{G}\,\,\hat{\vec{q}}_{1}^{\;-4}|\bar{B}B\rangle=\langle
A\bar{A}|n\nu\rangle\langle n\nu|\hat{G}|\hat{\vec{q}}^{\,\,-4}|\bar
{B}B\rangle
\]%
\[
=\sum_{n}\int_{-\infty}^{\infty}d\nu\,\Phi_{A}^{\ast}(n,\nu)\left(  \Phi
_{B}(n,\nu)+\frac{\bar{\alpha}^{2}\beta i}{4}[\Phi_{B}(n,\nu)\chi\left(
n,\nu\right)  ]^{\prime}Y\right)
\]%
\begin{equation}
\times e^{\bar{\alpha}\chi\left(  n,\nu\right)  (Y-Y_{0})+\frac{\bar{\alpha
}^{2}}{4}\delta\left(  n,\nu\right)  Y+\frac{\bar{\alpha}^{3}\beta i}{8}%
\chi\left(  n,\nu\right)  \chi^{\prime}\left(  n,\nu\right)  Y^{2}%
}.\label{discontinuityResult}%
\end{equation}
It is also interesting to consider the next term in series
(\ref{F-1decomposition}), though it is beyond the NLO accuracy. It reads%
\begin{equation}
\frac{\chi\left(  n,\nu\right)  ^{2}\chi^{\prime\prime}\left(  n,\nu\right)
+\chi\left(  n,\nu\right)  \chi^{\prime}\left(  n,\nu\right)  ^{2}}{3!}\left(
\frac{\bar{\alpha}^{2}\beta i}{4}Y\right)  ^{3}%
\end{equation}
since%
\begin{equation}
\frac{d^{3}F^{-1}(x)}{d^{3}x}|_{x=F\left(  t\right)  }=\left[  \chi\left(
n,t\right)  \right]  ^{2}\chi^{\prime\prime}\left(  n,t\right)  +\chi\left(
n,t\right)  \left[  \chi^{\prime}\left(  n,t\right)  \right]  ^{2}.
\end{equation}
Therefore the solution\ will have the form%
\[
\langle A\bar{A}|\hat{G}\,\,\hat{\vec{q}}_{1}^{\;-4}|\bar{B}B\rangle=\sum
_{n}\int_{-\infty}^{\infty}dv\,\Phi_{A}^{\ast}(n,\nu)\left(  \Phi_{B}%
(n,\nu)+\frac{\bar{\alpha}^{2}\beta i}{4}[\Phi_{B}(n,\nu)\chi\left(
n,\nu\right)  ]^{\prime}Y\right)
\]%
\begin{equation}
\times e^{\bar{\alpha}\chi\left(  n,\nu\right)  (Y-Y_{0})+\frac{\bar{\alpha
}^{2}}{4}\delta\left(  n,\nu\right)  Y+\frac{\bar{\alpha}^{3}\beta i}{8}%
\chi\left(  n,\nu\right)  \chi^{\prime}\left(  n,\nu\right)  Y^{2}-\frac
{\bar{\alpha}^{5}\beta^{2}Y^{3}}{3!4^{2}}\left(  \chi\left(  n,\nu\right)
^{2}\chi^{\prime\prime}\left(  n,\nu\right)  +\chi\left(  n,\nu\right)
\chi^{\prime}\left(  n,\nu\right)  ^{2}\right)  }.
\end{equation}
If one integrates with respect to $\nu$ treating the terms $\sim\bar{\alpha
}^{2}Y,\bar{\alpha}^{3}Y^{2},\bar{\alpha}^{5}Y^{3}$ as small and takes only
the contribution of the saddle point of $\chi\left(  0,0\right)  ,$ one gets%
\[
\langle A\bar{A}|\hat{G}\,\hat{\vec{q}}_{1}^{\;-4}|\bar{B}B\rangle\simeq
\,\Phi_{A}^{\ast}(0,0)\left(  \Phi_{B}(0,0)+\frac{\bar{\alpha}^{2}\beta i}%
{4}\Phi_{B}^{\prime}(0,0)\chi\left(  0,0\right)  Y\right)
\]%
\begin{equation}
\times\sqrt{\frac{2\pi}{\bar{\alpha}\left\vert \chi^{\prime\prime}\left(
0,0\right)  \right\vert Y}}e^{\bar{\alpha}\chi\left(  0,0\right)
(Y-Y_{0})+\frac{\bar{\alpha}^{2}}{4}\delta\left(  0,0\right)  Y-\frac
{\bar{\alpha}^{5}\beta^{2}Y^{3}}{3!4^{2}}\chi\left(  0,0\right)  ^{2}%
\chi^{\prime\prime}\left(  0,0\right)  }.
\end{equation}
Here the term\ $\sim\bar{\alpha}^{5}\beta^{2}Y^{3}$ has the same coefficient
as in \cite{Kovchegov:1998ae}, and the term $\sim\bar{\alpha}^{3}\beta
i\chi^{\prime}\left(  n,\nu\right)  $ is zero in the saddle point of $\chi$.
The latter is the reason why it did not appear in \cite{Kovchegov:1998ae}.

\section{Conclusion}

This paper presents the solution to the NLO forward BFKL equation in the space
of its Born eigenfunctions. In this space the equation can be written as a
prtial derivative equation. Its genral solution is given in (\ref{gensol}).
The solution obeying the boundary condition at $Y=Y_{0}$ can be obtained in
terms of inverse functions. However, in NLO these functions must be Taylor
expanded, which gives solution of the forward equation (\ref{<nu|BB>}) and
s-channel discontinuity (\ref{discontinuityResult}) with the NLO accuracy. The
exponent of the solution has a term $\sim\bar{\alpha}^{3}\beta i\chi^{\prime
}\left(  n,\nu\right)  Y^{2},$ which vanishes in the saddle point of the LO
eigenvalue. However, if one goes beyond the LO saddle point approximation it
is to be taken into account. \acknowledgments The study was supported by the
Ministry of education of the Russian Federation projects 14.B37.21.1181 and
8408 and by the Russian Fund for Basic Research grants 12-02-31086,
13-02-01023 and 12-02-33140 and by president grant MK-525.2013.2. I thank
prof. V.S. Fadin for proposing this work and prof. G. Salam for a helpful
discussion. I also thank Nordita for the invitation to and the hospitality and
support during the program \textquotedblleft Beyond the LHC\textquotedblright%
\ where this work was completed as well as the Dynasty foundation for the
financial support, which allowed me to attend this program.

\end{document}